\documentclass[10pt,twocolumn,letterpaper]{article}

\usepackage{cvpr}              % To produce the CAMERA-READY version
\usepackage{graphicx}
\usepackage[accsupp]{axessibility}
\usepackage{amsmath}
\usepackage{amssymb}
\usepackage{booktabs}
\usepackage{indentfirst}
\usepackage{algorithm} %format of the algorithm 
\usepackage{algorithmic} %format of the algorithm 
\usepackage{multirow} %multirow for format of table 
\usepackage{url}
\usepackage{makecell,rotating,multirow,diagbox}
\usepackage{verbatim}
\usepackage{caption}
\usepackage[pagebackref,breaklinks,colorlinks]{hyperref}

\usepackage[capitalize]{cleveref}
\crefname{section}{Sec.}{Secs.}
\Crefname{section}{Section}{Sections}
\Crefname{table}{Table}{Tables}
\crefname{table}{Tab.}{Tabs.}

\newcommand{\bfred}[1]{{\color{red}\textbf{#1}}}

\def\cvprPaperID{44} % *** Enter the CVPR Paper ID here
\def\confName{CVPR}
\def\confYear{2022}

\begin{document}

%%%%%%%%% TITLE - PLEASE UPDATE
\title{Slimmable Video Codec}

\author{Zhaocheng Liu$^{1}$, Luis Herranz$^{2}$\thanks{L.H. acknowledges the support of the Ramón y Cajal grant RYC2019-027020-I (MICINN, Spain).}, Fei Yang$^{2}$, Saiping Zhang$^{4}$, Shuai Wan$^{1}$, Marta Mrak$^{3}$ \\and Marc Górriz Blanch$^{3}$\\
$^1$ School of Electronics and Information, Northwestern Polytechnical University, Xi’an, China\\
$^2$ Computer Vision Center, Universitat Autonoma de Barcelona, 08193 Barcelona, Spain\\
$^3$ BBC Research \& Development, The Lighthouse, White City Place, 201 Wood Lane, London, UK\\
$^4$ State Key Laboratory of Integrated Services Networks, Xidian University, Xi’an, China\\
{\tt\small liuzhaocheng@mail.nwpu.edu.cn}
}
\maketitle

%%%%%%%%% ABSTRACT
\begin{abstract}
Neural video compression has emerged as a novel paradigm combining trainable multilayer neural networks and machine learning, achieving competitive rate-distortion (RD) performances, but still remaining impractical due to heavy neural architectures, with large memory and computational demands. In addition, models are usually optimized for a single RD tradeoff. Recent slimmable image codecs can dynamically adjust their model capacity to gracefully reduce the memory and computation requirements, without harming RD performance. In this paper we propose a slimmable video codec (SlimVC), by integrating a slimmable temporal entropy model in a slimmable autoencoder. Despite a significantly more complex architecture, we show that slimming remains a powerful mechanism to control rate, memory footprint, computational cost and latency, all being important requirements for practical video compression.
\end{abstract}

%%%%%%%%% BODY TEXT
\section{Introduction}
\label{sec:intro}

During the last two decades, video has become the dominant form of communication of the digital society. This has led to an explosive growth where video content accounts for more than 80\% of global data traffic. The \textit{basic (lossy) video compression} objective consists of transmitting as few bits as possible (i.e. minimize \textit{rate}) while representing the input sequence at a certain level of fidelity (i.e. \textit{distortion}). Video is now consumed using heterogeneous devices ranging from TV sets to smartphones. Furthermore, real-time video conferencing has become a household technology, pervasive in work and educational environments. These practical scenarios imposes additional constraints to the design of video codec in practice, such as dynamically controllable rate, low computational and memory footprint, and low latency. Together with the previous rate and distortion objectives, they conform the more challenging problem of \textit{practical video compression}.

In parallel, the deep learning revolution has motivated a new compression paradigm based on parametric encoders and decoders implemented as deep neural networks which are optimized with data. This compression approach has been applied successfully first in images~\cite{2016arXiv161101704B,2018arXiv180201436B,2018arXiv180902736M} and then videos~\cite{DVC,HLVC}. This paradigm contrasts with the traditional hybrid video coding paradigm, based on block-based linear transforms and carefully engineered coding tools (e.g. H.264/AVC, H.265/HEVC). Focusing on improving rate-distortion performance, most neural image and video codecs are impractical, since require heavy and complex networks. Practical aspects have been always carefully considered in the design of traditional codecs. In contrast to previous works, our paper focuses chiefly on those practical constraints, proposing a lightweight and flexible design for practical neural video compression.

Our design is based on a slimmable autoencoder augmented with a slimmable temporal entropy model. This design is motivated by two recent works. Motivated by the empirical observation that lower rates do not require the use of full capacity, Yang \textit{et al.}~\cite{9578334}  proposed the slimmable compressive autoencoder (SlimCAE) architecture, where the slimming becomes a flexible mechanism to both vary the rate-distortion tradeoff and control the complexity. However, extending SlimCAE to video by including temporal prediction is not trivial, since most designs require additional modules to estimate and compensate motion (e.g. optical flow nets, motion compensation nets). Slimmable designs of such modules are not straightforward, nor the potential interplay with other elements in the compression framework. Recently, Sun \textit{et al.}~\cite{STEM} proposed spatiotemporal entropy model (STEM), a motion-free framework where temporal prediction is performed directly in the entropy model without any motion estimation nor compensation. In our framework we adopt part of STEM's entropy model and propose a slimmable version, thus having a fully slimmable codec.

In summary, this work contributes with a novel slimmable video codec (SlimVC) designed to address practical challenges in the neural video compression paradigm, via a simple slimming mechanism. Experiments show that our slimmable model can effectively exploit temporal redundancy without a significant drop in RD performance compared to that of independent models. 

%-------------------------------------------------------------------------
\section{SlimCAE and STEM}

\subsection{Slimmable compressive autoencoder}
Neural image codecs are implemented typically as compressive autoencoders (CAEs)~\cite{theis2017lossy,2016arXiv161101704B}, consisting of autoencoders augmented with quantization and entropy coding. The encoder $\mathbf{z}=f\left(\mathbf{x};\theta\right)$ parametrized by $\theta$
transforms the input image $\mathbf{x}$ into a latent representation $\mathbf{z}$, which is then quantized as $\mathbf{q}=Q\left(\mathbf{z}\right)$ and the entropy encoder maps it to the bitstream $\mathbf{b}$. In the decoder, $\mathbf{b}$ is mapped back to the reconstructed latent representation $\hat{\mathbf{z}}$, and the decoder parametrized by $\phi$ recovers the reconstructed image $\hat{\mathbf{x}}=g\left(\hat{\mathbf{z}};\phi\right)$. During training, quantization is replaced by a differentiable proxy are used (additive uniform noise, in our case) and entropy coding is bypassed and the rate is approximated by the entropy of the latent representation. This requires a model of the probability distribution parametrized by $\nu$. This model, usually refer to as entropy model, has been the source of many improvements in RD performance, by including hyperpriors~\cite{2018arXiv180201436B} and autoregressive models~\cite{2018arXiv180902736M}.

CAEs are typically trained by minimizing a RD objective
\begin{equation}
    \mathcal{L}\left(\theta,\phi,\nu;\mathcal{X},\lambda\right)=D\left(\theta,\phi,\nu;\mathcal{X}\right)+\lambda R\left(\theta;\mathcal{X}\right),
    \label{eq:CAE_loss}
\end{equation}
where $\mathcal{X}$ is the set of training images, $\lambda$ is the tradeoff between the rate $R$ of the latent representations and distortion $D$ between input and reconstructed images, averaged over $\mathcal{X}$.

An slimmable compressive autoencoder (SlimCAE)~\cite{9578334} is a CAE whose layers are slimmable. The slimmable layers can discard part of the parameters while still performing a valid operation. This results in less expressivenes, but also lower memory footprint and computation. The SlimCAE contains $K$ sub-models, each of which is determined by a set of parameters $\left(\theta^{(k)},\phi^{(k)},\nu^{(k)}\right)\!\in\!\Psi^{(k)}$, where $k\!=\!1,...,K$. The parameters of the sub-model $k$ are a superset of the parameters in the sub-model $k-1$. Finally, the $K$ sub-models are trained jointly using a joint loss 
\begin{equation}
    \mathcal{L}\left(\Psi;\mathcal{X}\right)=\sum_{1}^K\mathcal{L}^{(k)}\left(\theta^{(k)},\phi^{(k)};\mathcal{X}\right).
    \label{eq:slimCAEloss}
\end{equation}
In~\cite{9578334}, the authors showed that if the set of $\lambda^{(k)}$ are determined properly for the specific sub-modules, SlimCAE can achieve roughly the same RD performance of independent models optimized for single fixed $\lambda$s.

\subsection{Spatiotemporal entropy model }
\begin{figure*}[h]
	\centering
	\includegraphics[width=\textwidth]{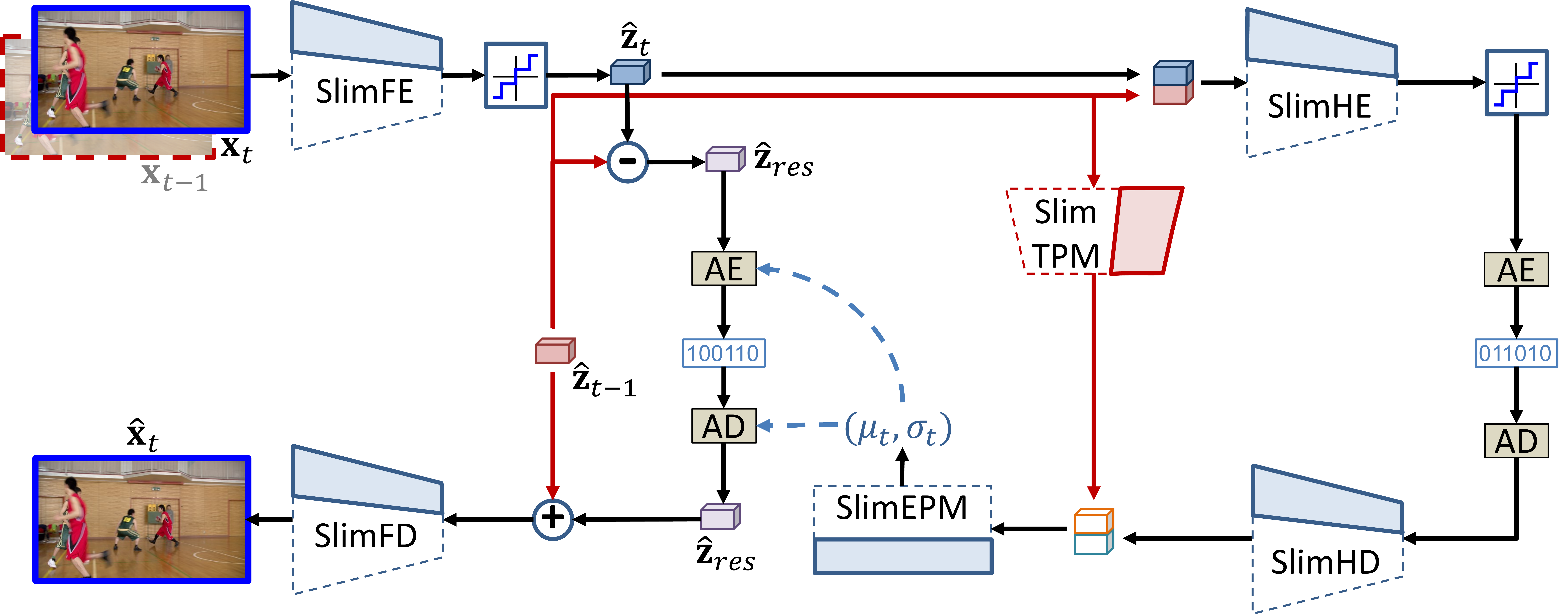}
	\caption{Slimmable video codec framework. In the slimmable modules (SlimFE, SlimFD, SlimHE, SlimHD, SlimTPM and SlimEPM), dashed lines represent the full capacity of the module, and solid ones the specific capacity after slimming to a particular operating point.}
	\label{FigureOne}
\end{figure*}

Sun \textit{et al.}~\cite{STEM} proposed a motion-free video compression method observing that inter-frame redundacy can be exploited efficiently in the entropy module via a spatiotemporal entropy model (STEM)\cite{STEM} without requiring motion estimation. In this model, the hyperencoder (HE) of the hyperprior receives the latent representations $\hat{\mathbf{z}}_t$ and $\hat{\mathbf{z}}_{t-1}$ of both the current frame and the previous one, allowing it to exploit temporal redundancy, reducing the rate of the side information received by the hyperdecoder (HD). In addition, only the residual latent $\hat{\mathbf{z}}_{res}=\hat{\mathbf{z}}_t-\hat{\mathbf{z}}_{t-1}$ is transmitted in the bitstream. In order to obtain more accurate distribution models, while further exploiting spatial and temporal redundancy, STEM includes a spatial prior module (SPM) and a temporal prior module (TPM), together with an entropy parameters module (EPM) that fuses the information and predicts the actual distribution parameters at time $t$.

SPM is an autoregressive PixelCNN-like network, and provides a relatively minor gain in RD performance at significant increased computational cost and particularly, two orders of magnitud increase in latency (from tenths to tens of seconds, both reported by \cite{STEM} and verified in our implementation). For these practical reasons, we chose not to include SPM in our framework.

\section{Fully slimmable framework}
The proposed framework is shown in Fig.~\ref{FigureOne}, where all trainable modules are designed to be slimmable\footnote{We use switchable GDNs .}, including both the feature autoencoder (i.e. SlimFE, SlimFD) and the entropy model (i.e. SlimHE, SlimHD, SlimTPM, SlimEPM). For simplicity, we assume uniform slimming, that is, the width (i.e. number of channels) in every slimmable layer is slimmed by the same factor (we use the same factors as in SlimCAE~\cite{9578334}, i.e. [0.25,0.375,0.5,0.75,1]). Table~\ref{architecture_details} provides more details about the architecture of the slimmable modules. SlimVC is trained in two stages. First, we train it as an image-based SlimCAE with hyperprior. Then we discard the hyperprior, and add the remaining modules of SlimVC (note that SlimCAE's hyperprior is image-based, while SlimHE and SlimHD have distinct architectures and designed for pairs of frames). Then we fix SlimFE and SlimFD, and train the remaining slimmable modules.

\begin{table*}[h]\footnotesize
	\centering
	\caption{Details of the slimmable modules in our implementation of SlimVC. Width factors: \bfred{0.25/0.375/0.5/0.75/1}. \textbf{SConv}/\textbf{SDeconv}: slimmable convolution/transposed convolution, \textbf{swGDN}/\textbf{swIGDN}: switchable GDN/IGDN, \textbf{LReLU}: Leaky ReLU.}
	\resizebox{\textwidth}{!}{
		\setlength{\tabcolsep}{0.5mm}{
			
			\begin{tabular}{c|c|c}
				\hline
				Module & Architecture & Params (millions)\\
				\hline
				SlimFE & \textbf{SConv}9x9s3\bfred{c48/72/96/144/192}-\textbf{swGDN}-\textbf{SConv}5x5s2\bfred{c48/72/96/144/192}-\textbf{swGDN}-\textbf{SConv}5x5s2\bfred{c48/72/96/144/192}-\textbf{swGDN} & \bfred{0.1/0.3/0.5/1.1/2}\\
				SlimFD & \textbf{swIGDN}-\textbf{SDeconv}5x5s2\bfred{c48/72/96/144/192}-\textbf{swIGDN}-\textbf{SDeconv}5x5s2\bfred{c48/72/96/144/192}-\textbf{swIGDN}-\textbf{SDeconv}9x9s3\bfred{c48/72/96/144/192} & \bfred{0.1/0.3/0.5/1.1/2}\\
				SlimHE &\textbf{SConv}3x3s1\bfred{c64/96/128/192/256}-\textbf{LReLU}-\textbf{SConv}5x5s2\bfred{c64/96/128/192/256}-\textbf{LReLU}-\textbf{SConv}5x5s2\bfred{c64/96/128/192/256} &\bfred{0.7/1.2/1.7/2.8/4.2}\\
				SlimHD &\textbf{SConv}5x5s2\bfred{c64/96/128/192/256}-\textbf{LReLU}-\textbf{SConv}5x5s2\bfred{c64/96/128/192/256}-\textbf{LReLU}-\textbf{SConv}3x3s1\bfred{c160/240/320/480/640} &\bfred{0.9/1.4/2.0/3.3/4.8}\\
				SlimTPM &\textbf{SConv}5x5s1\bfred{c107/160/213/320/426}-\textbf{LReLU}-\textbf{SConv}5x5s1\bfred{c133/200/267/400/533}-\textbf{LReLU}-\textbf{SConv}5x5s1\bfred{c160/240/320/480/640} &\bfred{3.0/4.8/6.7/11.1/16.2}\\
				SlimEPM &\textbf{SConv}1x1s1\bfred{c400/600/800/1200/1600}-\textbf{LReLU}-\textbf{SConv}1x1s1\bfred{c320/480/640/960/1280}-\textbf{LReLU}-\textbf{SConv}1x1s1\bfred{c96/144/192/288/384} &\bfred{0.8/1.2/1.8/3.0/4.6}\\
				\hline
			\end{tabular}
			\label{architecture_details}}
	}
\end{table*}

\section{Experiments}
\subsection{Experimental settings}
\paragraph{Datasets and training details} We use Open Images \cite{openimages} and CLIC as training datasets \cite{CLIC} during the first training stage, with random $256\times 256$ crops and a batch size of 16 crops. For the second stage, we use small sequences from the Vimeo-90k dataset \cite{2017arXiv171109078X}, in $256\times 256$ pixel crops and a batch size of 32 crops. The model has five RD operating points (i.e. [0.25,0.375,0.5,0.75,1], as mentioned earlier). We use a learning rate of 5e-5, and mean square error (MSE) as distortion metric.

\paragraph{Methods} \textbf{SlimVC (GOP=$N$)}: the proposed approach after the second stage of training with a group of pictures of size $N$.  \textbf{SlimVC (intra-only)}: is the codec resulting from the first stage without exploiting temporal redundancy. \textbf{Independent VCs (GOP=$N$)}: uses the same architecture of SlimVC but with a single width, so each RD point corresponds to a different model trained independently for that specific RD tradeoff. For comparison we also include H.264, STEM\cite{STEM} and DVC~\cite{DVC}. Note that DVC is significantly more complex, and uses  motion estimation and compensation with temporal prediction in the pixel domain.

\subsection{Rate-distortion}
\begin{figure}[h]
	\centering
	\includegraphics[scale=0.65]{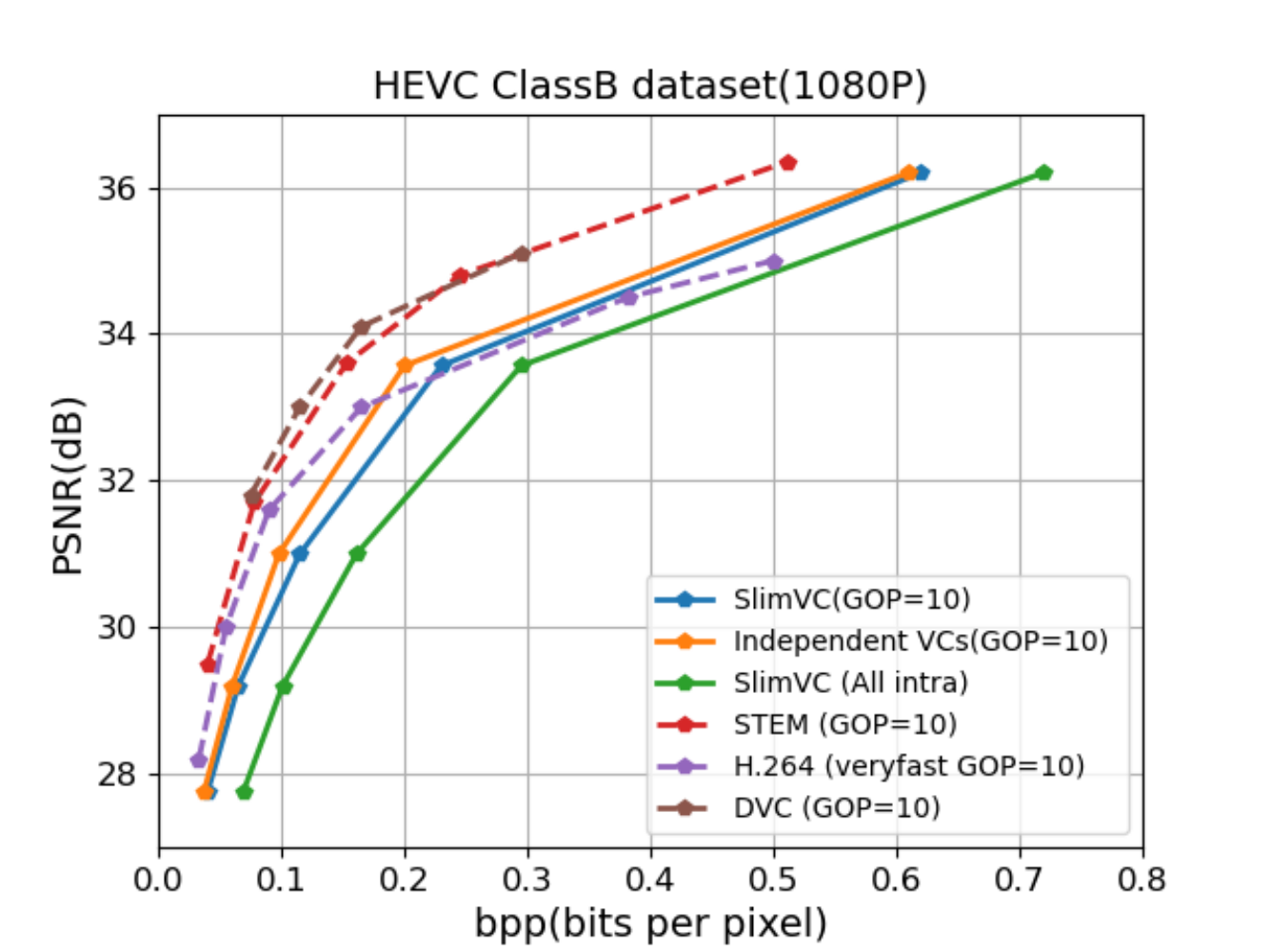}
	%\hspace{-10mm}
	\includegraphics[scale=0.65]{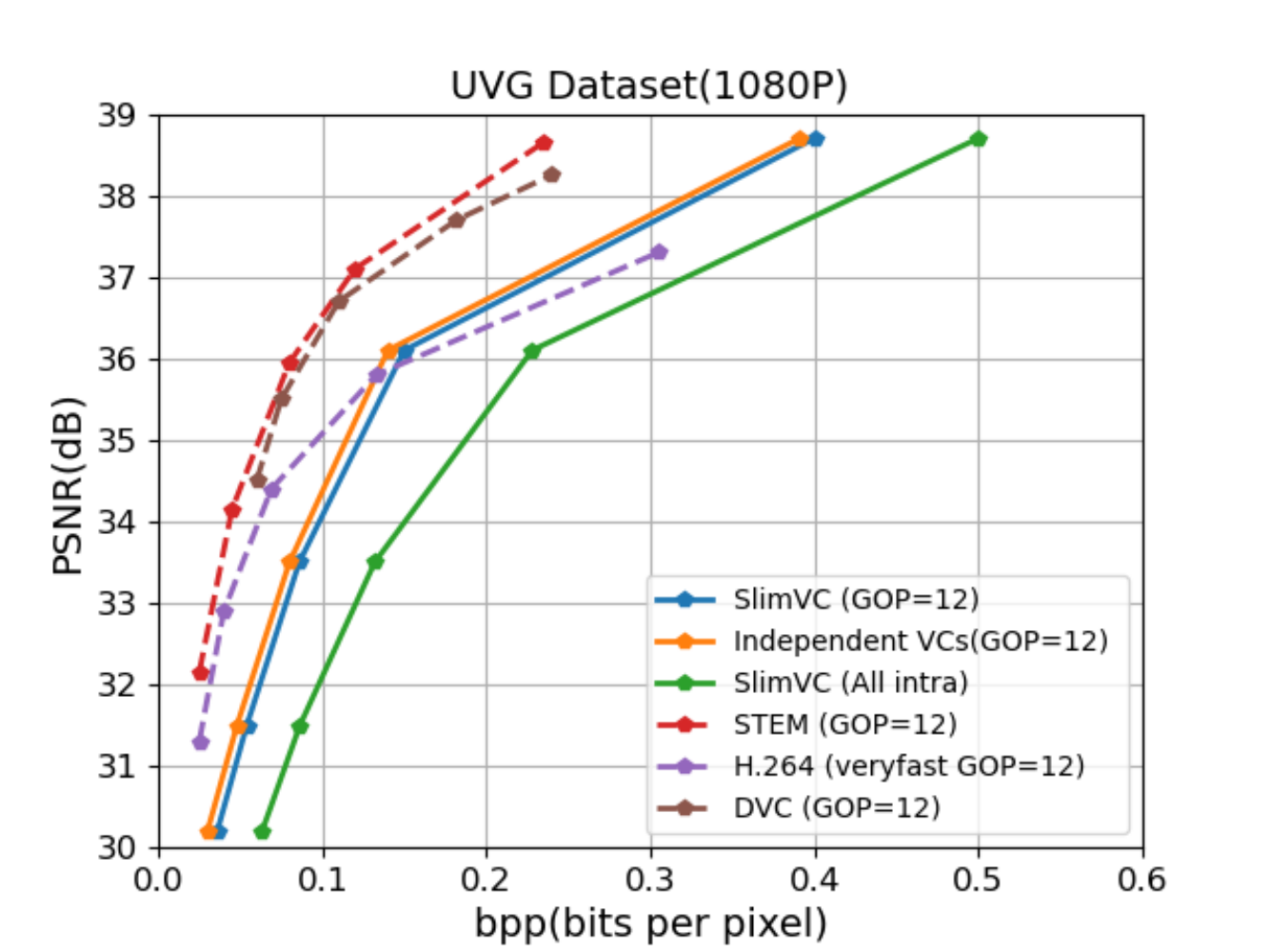}
	\caption{Rate-distortion performance on the HEVC Class B dataset (top) and UVG dataset (bottom).}
	\label{RD}
\end{figure}

We compressed the first 100 frames of HEVC Class B sequences \cite{6316136} and the Ultra Video Group test sequences \cite{UVG} with a GOP size of 10 and 12 pictures, respectively. The RD performances of the different methods are shown in Fig.\ref{RD}\footnote{We included the RD curve of STEM from \cite{STEM} for reference, but note that the architectures are not comparable: the implementation of STEM in \cite{STEM} uses encoders and decoders with four convolutional layers, while we use three, and their entropy model leverages an autoregressive context model and an SPM, which are not used in our case.}. The proposed SlimVC has a RD performance very close to that of independently trained VCs, thus showing the benefit of SlimVC in terms of providing variable rate with one single model. Comparing with SlimVC (all intra), we can see that the slimmable temporal entropy model and the second stage are effective in consistently reducing the rate at all RD points (SlimVC curves are shifted towards the left). RD performance is comparable to that of H.264, and remains below that of DVC, which is significantly more complex and lacks the flexibility of SlimVC (see next section). Besides, the design of SlimVC has still considerable room form improvement of RD performance.

\subsubsection{Memory and computational efficiency}
We measured the efficiency of SlimVC and other baselines in terms of computational cost (in floating point operations, FLOPs) and memory footprint (in MB) when processing 1080P input sequences (i.e., 1920×1080×3). Table~\ref{computation} shows that SlimVC requires significantly less computations than the other video baselines, especially in lower rates where the slimmable design is able to avoid most of the computations, leading to very significant speedups (up to 20x for low rates).

\begin{figure}[h]
	\centering
	\includegraphics[scale=0.375]{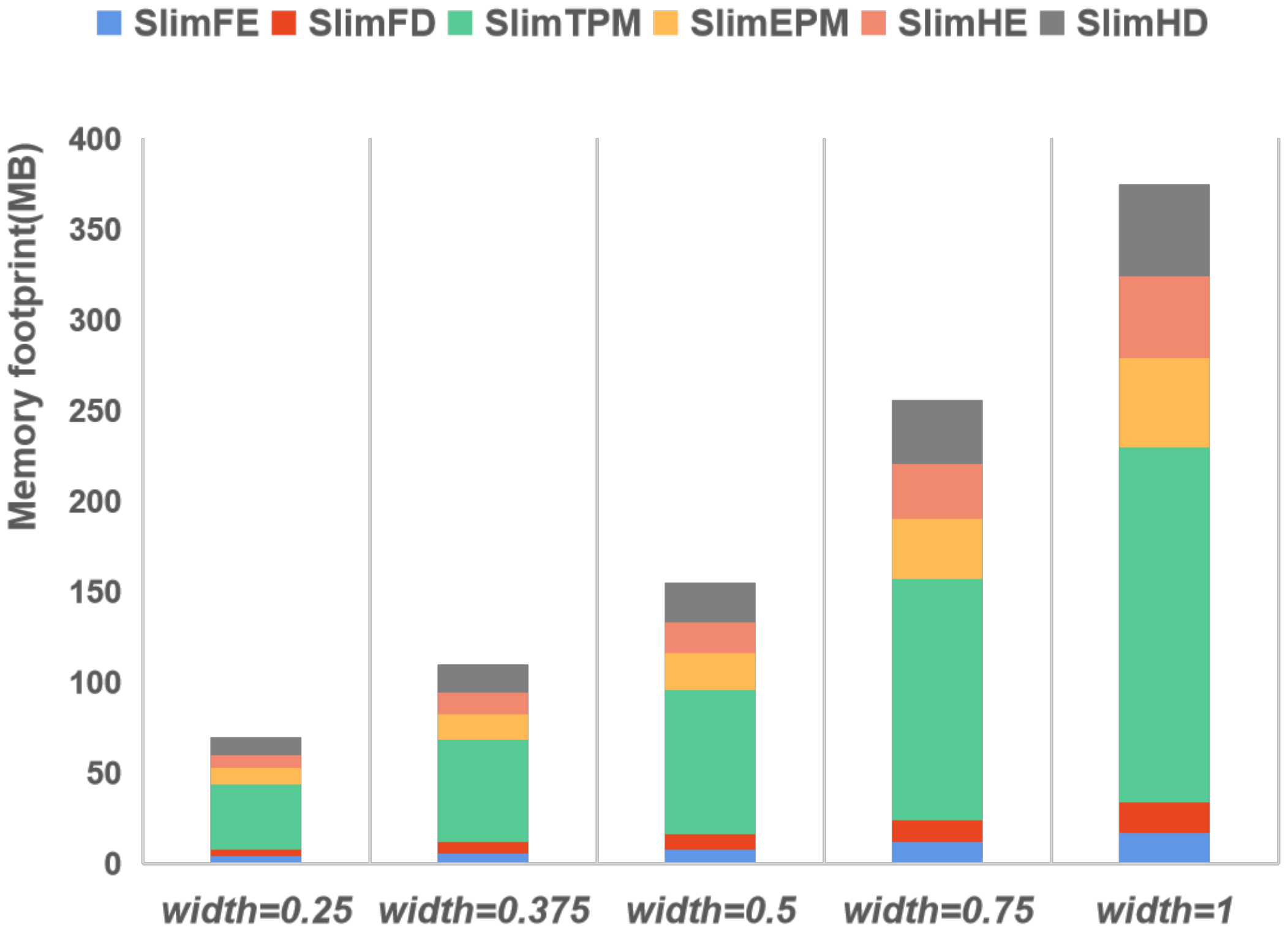}
	
	\caption{Memory footprint of SlimVC for different widths (and correspondingly, bit rates).}
	\label{memory}
\end{figure}

Fig.\ref{memory} shows the detailed memory footprint of SlimVC and the different modules for the different widths. It shows that SlimVC is a lightweight method, whose memory footprint can be gracefully adjusted depending on the rate needs. In contrast to SlimCAE, where the feature encoder and decoder were the main bottlenecks in terms of memory and computation, in SlimVC the most critical modules in this regard are those related to entropy modeling. In particular, the temporal prediction module is the heaviest module of the codec.

\begin{table}[h]\footnotesize
	\centering
	\caption{Computational cost (GFLOPs) for the different methods for 1080P sequences.}
	\setlength{\tabcolsep}{0.4mm}{
		\begin{tabular}{|c|c|c|ccccc|}
			\hline \multicolumn{3}{|c|}{Methods} & Low rate & $\Rightarrow$ & Medium rate & $\Rightarrow$ & High rate \\
			\hline
			\multirow{9}*{\rotatebox{90}{Encoding}} & \multirow{5}*{\shortstack{SlimVC}} & SlimFE/FD & 9.6 & 18.5 & 34 & 56 & 90  \\
			& & SlimTPM & 42.4 & 68 & 95.7 & 160 & 232.5  \\
			& & SlimEPM & 11 & 17.5 & 25.7 & 43.9 & 66  \\
			& & SlimHE/HD & 10 & 15 & 20.7 & 33.3 & 47.6 \\
			& & Total & \textbf{73} & \textbf{119} & \textbf{176} & \textbf{293} & \textbf{436} \\\cline{2-8}
			
			& \multicolumn{2}{c|}{Indep. VCs} & \textbf{73} & \textbf{119} & \textbf{176} & \textbf{293} & \textbf{436} \\
			\cline{2-8} &\multicolumn{2}{c|}{STEM} & 643 & 643 & 643 & 643 & 643 \\\cline{2-8}
			&\multicolumn{2}{c|}{STEM w/o SPM} & 613 & 613 & 613 & 613 & 613\\\cline{2-8}
			& \multicolumn{2}{c|}{DVC} & 3074 & 3074 & 3074 & 3074 & 3074\\\cline{2-8}
			
			\hline \multirow{9}*{\rotatebox{90}{Decoding}} &  \multirow{5}*{\shortstack{SlimVC}} & SlimFE/FD & 9.6 & 18.5 & 34 & 56 & 90  \\
			& & SlimTPM & 42.4 & 68 & 95.7 & 160 & 232.5  \\
			& & SlimEPM & 11 & 17.5 & 25.7 & 43.9 & 66  \\
			& & SlimHD & 6 & 9 & 12.2 & 19.4 & 27.4  \\
			& & Total & \textbf{69} & \textbf{113} & \textbf{168} & \textbf{279} & \textbf{416}  \\\cline{2-8}
			& \multicolumn{2}{c|}{Indep. VCs} & \textbf{69} & \textbf{113} & \textbf{168} & \textbf{279} & \textbf{416}  \\\cline{2-8}			
			& \multicolumn{2}{c|}{STEM} & 1509 & 1509 & 1509 & 1509 & 1509\\\cline{2-8}
			&\multicolumn{2}{c|}{STEM w/o SPM} & 1479 & 1479 & 1479 & 1479 & 1479\\\cline{2-8}
			& \multicolumn{2}{c|}{DVC} & 1434 & 1434 & 1434 & 1434 & 1434\\\cline{2-8}
			\hline
		\end{tabular}
		\label{computation}}
\end{table}

\section{Conclusion}
Motivated by some practical limitations of current neural video codecs, we propose slimmable video codec (SlimVC), a novel adaptive architecture based on slimmable modules that can provide significant savings in memory and computational costs for low and mid rates, together with variable rate control with one single video model. While SlimCAE showed that slimmable codecs are promising approaches for practical neural image compression, SlimVC further advances this potential for the case of practical neural video compression.

%%%%%%%%% REFERENCES
{\small
\bibliographystyle{ieee_fullname}
\bibliography{egbib}
}

\end{document}